\documentstyle[11pt,paspconf,epsf]{article}
  \input{epsf.sty}

\newcommand\th{\thinspace} 
\newcommand\kms{\ifmmode{\rm km\th s^{-1}}\else km\th s$^{-1}$\fi} 
\newcommand\ms{\ifmmode{\rm m\th s^{-1}}\else m\th s$^{-1}$\fi} 
\newcommand\Mo{\ifmmode{M_{\odot}}\else $M_{\odot}$\fi}  
\newcommand\MJ{\ifmmode{M_{Jup}}\else $M_{Jup}$\fi}

\begin{document} 

\title{COMPARISON BETWEEN EXTRASOLAR PLANETS AND LOW-MASS
SECONDARIES}

\author{Tsevi Mazeh and Shay Zucker}
\affil{School of Physics and Astronomy, Raymond and Beverly Sackler  
Faculty of Exact Sciences, Tel Aviv University, Tel Aviv, Israel\\
e-mail: mazeh, shay@wise.tau.ac.il}

\begin{abstract} 

This paper compares the statistical features of the sample of discovered
extrasolar planets with those of the secondaries in nearby spectroscopic
binaries, in order to enable us to distinguish between the two
populations.  Based on 32 planet candidates discovered until March 2000,
we find that their eccentricity and period distribution are surprisingly
similar to those of the binary population, while their mass distribution
is remarkably different. The mass distributions definitely support the
idea of two distinct populations, suggesting the planet candidates are
indeed extrasolar planets. The transition between the two populations
probably occurs at 10--30 Jupiter masses. We point out a possible
negative correlation between the orbital period of the planets and the
metallicity of their parent stars, which holds only for periods less
than about 100 days. These short-period systems are characterized by
circular or almost circular orbits.

\end{abstract}

\section{Introduction}

In the last few years we are witnessing a burst of discoveries of
candidates for extrasolar planets (for a recent review see Marcy,
Cochran, \& Mayor 2000, hereafter MCM). These 'planet candidates' were
discovered by detecting small periodic radial-velocity modulations of
their parent stars, which indicate the existence of unseen companions.
The identification of the companions as planet candidates is based
solely on their inferred minimum masses, which are of the order of a
Jupiter mass (=\MJ).

This identification is based on the commonly accepted notion that planet
masses are substantially smaller than those of stars. Some works go even
further and {\it define} a planet as an object with mass smaller than 13
\MJ --- the minimum mass needed to ignite deuterium in its core.  This
definition goes hand in hand with the definition of a brown dwarf as an
object that does not burn hydrogen in its core, and therefore has a mass
less than about 80 \MJ.  The mass-based definition of a planet became so
popular, that some astronomers used it without requiring a planet to
orbit another star. This is reflected, for example, in the title of a
recent paper {\it A Population of Very Young Brown Dwarfs and
Free-Floating Planets in Orion} by Lucas \& Roche (2000).

Obviously, we can arbitrarily adopt any definition for any
term. However, we usually expect a good definition to carry along any
previous understandings of the term. Planets were conceived in the past
as objects orbiting the Sun --- a feature completely missing from the
mass definition of a planet. Actually, one of the first works to suggest
this distinction was that of Burrows et al. (1997), who emphasized that
this distinction was arbitrary for ``parsing by eye the information in
the(ir detailed) figures'', and they did not ``advocate abandoning the
definition based on origin''.

We therefore suggest a somewhat hypothetical, open-ended definition,
which is based on two notions. The first one is indeed that planets must
orbit their parent stars. This immediately raises the question of how to
distinguish between planets and low-mass secondaries in binary systems.
We therefore suggest that the definition includes a second notion, based
on the seemingly accepted paradigm that planets, including giant planets
like Jupiter, were formed differently than stars.  The present picture
is that planets were probably formed by coagulation of smaller, possibly
rocky, bodies,
while stars were probably formed by some kind of fragmentation of larger
bodies. In other words, planets were formed by small bodies that grew
larger, while stars, binary included, were formed by fragmentation of
large bodies into smaller objects.  Therefore, according to the proposed
definition, a planet is a low-mass object formed differently than a
star, orbiting a much larger star. Note that this definition does not
include {\it a priory} the planet mass, a specific mass limit in
particular.

The formation notion, if true, probably allows us to differentiate {\it
statistically} between giant planets and low-mass secondaries. A
population of planets might show some statistical features different
than the ones found in low-mass binaries, reflecting the difference in
formation history.  Finding such differences in observed samples can
verify the formation aspect of the definition, which, at this stage of
the research of extrasolar planets, is still only an assumption.
Therefore, this paper adopts a purely observational approach and checks
whether such distinguishing characteristics can be found, refraining
from any theoretical discussion of their origin.

We could expect, for example, that the distribution of orbital
eccentricities of giant planets and low-mass binaries will be
substantially different, because all the solar planets have nearly
circular orbits, while binaries do not (e.g., Mazeh, Mayor, \& Latham
1996). Or, we could expect the periods of planets to be longer than 10
years, like the giant planets in the solar system. Many studies of the
newly discovered planets showed that this is not the case (e.g.,
MCM). Moreover, following Heacox (1999) who based his analysis only
upon 15 binaries and a handful of planet candidates, we will show that
within some reasonable restrictions, the eccentricity and period
distributions of the two samples are surprisingly similar. On the other
hand, the mass distributions of the planet candidates and the low-mass
secondaries are well separated (e.g., Marcy \& Butler 1998), definitely
suggesting the existence of two populations. At the same time the very
different mass distributions validate the original notion of mass
difference. The border zone between the two populations could be
demarcated in the near future when more data are available.

\section{Eccentricity and Period Distribution}

We consider here the 32 planet candidates that were discovered until
March 2000 (Schneider 2000), comparing their orbital characteristics
with those of spectroscopic binaries. For the latter we use the results
of a very large radial-velocity study of the Carney \& Latham (1987)
high-proper-motion sample, which yielded 200 spectroscopic binaries
(Latham et al. 2000; Goldberg et al. 2000). Goldberg (2000) succeeded to
separate statistically between the binaries of the Galactic halo and
those coming from the disk. We consider in this section only the 59
single-lined spectroscopic binaries (=SB1s) of the Galactic disk.

\begin{figure}
\plotone{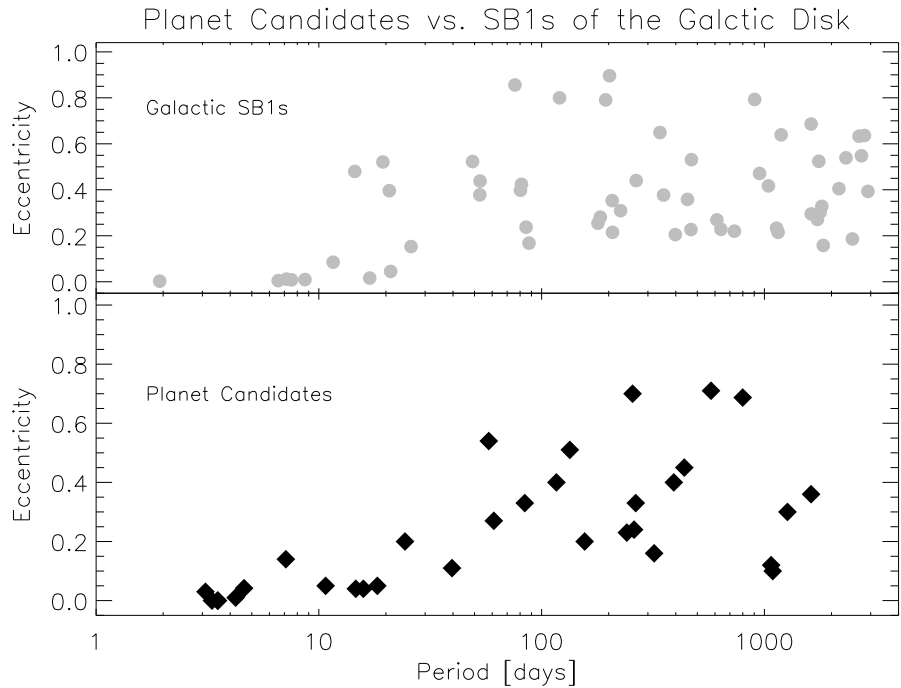}
\caption{The eccentricity as a function of the orbital period for the
Galactic disk SB1s and the planet candidates.}
\end{figure}

In Figure 1 we present the eccentricity-period relation for the two
samples. In both samples all the short-period binaries have circular or
almost circular orbits. Nevertheless, it seems as if there is some
subtle difference in the way this effect is revealed in the two
samples. At the upper panel there is a clear division; all binaries with
periods shorter than 10 days are practically circular, while binaries
with longer periods show considerable eccentricities. In the planet
candidate sample we find high eccentricities, above 0.2, only with
periods longer than 50 days. Orbits with shorter periods do not have
high eccentricities, although some of the orbits are not completely
circular. The difference between the two samples is certainly not well
established statistically, and we need many more planet candidates to
assess its reality.

Apparently, some tidal interaction circularized or nearly circularized
the orbits of the short period systems in both samples. Although the
exact mechanism is not yet clear (e.g., Zahn \& Bouchet 1989, Goldman
\& Mazeh 1991, Goodman \& Oh 1997), the cutoff shape of the SB1s is
well explained.  The eccentricity distribution of the planet candidates,
if different from that of the SB1s,
is more difficult to interpret, and has probably to do with the orbital
evolution that these systems have gone through (e.g., Wiedenschilling \&
Marzari 1996; Lin et al. 2000; Trilling 2000)

Next, we consider the eccentricity distribution of the uncircularized
orbits of the two samples. To do so we plot in Figure~2 the eccentricity
cumulative distribution of the two samples, with period longer than 10
days for the SB1 sample and with periods longer than 50 days for the
planet candidates. The result, first noted by Heacox (1999) who based
his analysis only upon 15 binaries, is astounding. The two populations
have practically the same distribution, at least for most of the
eccentricity range. Stepinski \& Black (2000, a poster paper in this
meeting) who used Heacox small sample of binaries, and Mayor \& Udry
(2000) came to similar conclusion.

\begin{figure}
\plotfiddle{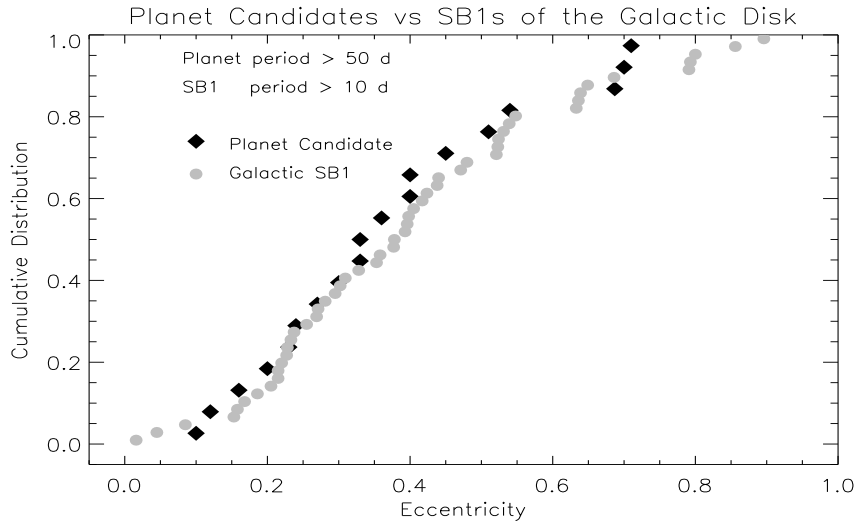}{6.7 truecm}{0}{70}{60}{-165}{-10}
\caption{The eccentricity cumulative distribution of the planet
candidates and the Galactic disk SB1s.}
\end{figure}

\begin{figure}
\plotfiddle{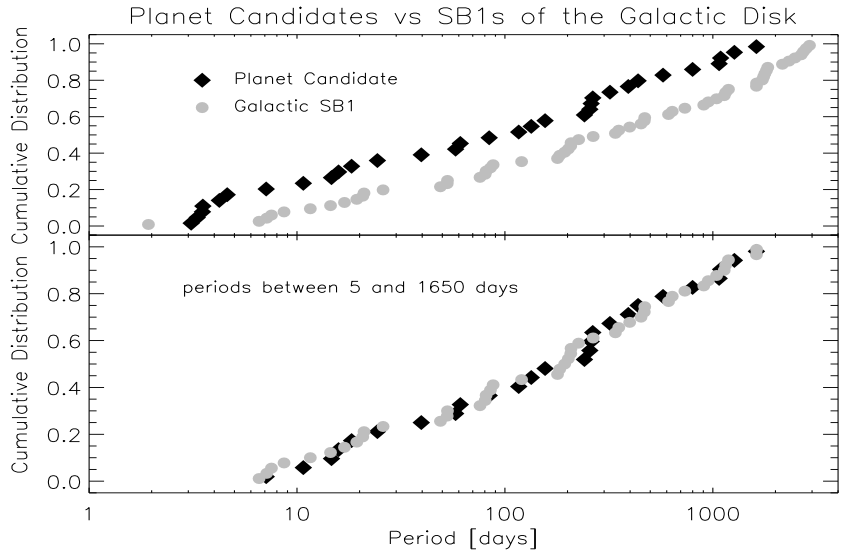}{6.7truecm}{0}{70}{60}{-165}{-5}
\caption{The period cumulative distribution of the planet candidates and
the Galactic disk SB1s. See text for the two panels.}
\end{figure}

In the upper panel of Figure 3 we compare the {\it period} distribution
of the two samples. The two distributions run parallel for most of the
period span, which indicates the same density distribution. To emphasize
this point we exclude binaries with period shorter than 7 days or longer
than 1650 days from both samples and plot in the lower panel the two
restricted distributions, which turn out to be the same. Moreover, as
noted already by Heacox (1999), the figure shows that the two
distributions, when plotted here on a logarithmic scale follow strictly
a straight line, which indicates flat density distributions on a
logarithmic scale. Stepinski \& Black (2000) came to
similar conclusion.

Any paradigm that assumes the two populations were formed
differently has to explain why their eccentricity as well as period
distributions are so much alike. One might wonder are they
really two separate populations. However, any such doubt can be put to
rest by considering the mass distribution, as is done in the next
section.  

\section{The Mass Distribution of the Planet Candidates and the Low-Mass
Secondaries}
 
\begin{figure}
\plotfiddle{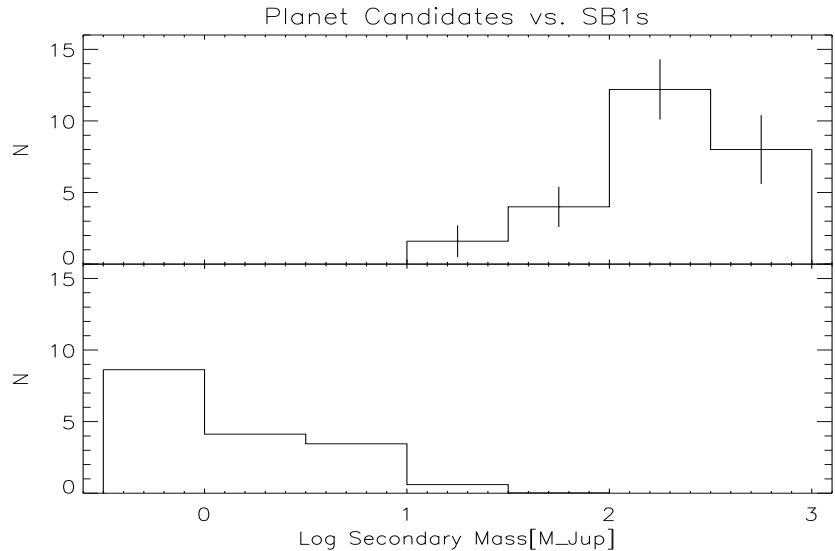}{6.3 truecm}{0}{70}{60}{-165}{-10}
\caption{Corrected histograms of the extrasolar planet candidates and
the low-mass secondaries of spectroscopic binaries. Upper panel for the
SB1s. Lower panel for the planet candidates. Both scaled for a sample of
200 stars}
\end{figure}

In Figure 4 we present two separate mass histograms, one of the SB1s and
the other of the planet candidates. We follow, with some slight
modifications, the derivations of Mazeh, Goldberg, \& Latham (1998,
hereafter MGL) and of Mazeh (1999a,b), which were done with a
substantially smaller sample of planet candidates. The detailed
derivation is explained in those papers. In short, the {\it observed}
histograms are modified so they take into account two effects. The first
has to do with the unknown inclination angles of most of the systems,
which allow the observer to derive only the minimum masses of the
unseen companions. The second effect reflects the fact that the
observers can not detect radial-velocity modulation with a too small
amplitude, either because of the low orbital inclination or because of
the smallness of the secondary mass. To be able to compare the two
distributions, which are spread over more than three orders of
magnitudes, it is important to present the data on a logarithmic scale.

The last two bins of the SB1s histogram, with masses larger than 100
\MJ, were derived from a subsample of the high-proper-motion binaries
(Latham et al. 1998), drawn from a sample of 420 primaries with masses
higher than 0.7 \Mo\ (Latham et al. 2000; Goldberg et al. 2000).  The
other two bins, between 10 and 100 \MJ, could not be derived from that
radial-velocity survey, because of lack of sensitivity.  The histogram
is based, instead, on the work of Mayor et al.\ (1997), who studied a
sample of 570 nearby K stars (see also Halbwachs, Mayor \& Udry
1998). Mayor et al.\ were kind enough to let MGL know that they have
found 5 additional binaries in that range.  The two sets of binaries
were drawn from samples of different sizes.  We therefore scaled both
pairs of bins to a sample size of 200 systems, the number of stars
included in the first phase of the planet search.

Note that the numbers of stars in the 10--30 \MJ\ bin, and even in the
30--100 \MJ\ one, are {\it statistically indistinguishable from zero},
consistent with the idea of a ``brown-dwarf desert'' (e.g., Halbwachs et
al. 2000). They used Hipparcos data and concluded that some of the
systems in these bins might have masses larger than 0.08 \Mo\ --- the
stellar border line, turning these 'brown-dwarf candidates' into stellar
companions. In any event, the secondary frequency, which rises when
moving from, say, 1000 \MJ ($\simeq 1M_{\odot}$), to 300 \MJ, drops down
very sharply and gets to very low values at the range of 30--100 \MJ.

The other histogram includes all the published and announced planet
candidates, until March 2000. This sample is obviously incomplete, as no
research group exhausted the planet discoveries in the sample they are
following. In particular, the stars with low-amplitude variations are
still being monitored so their periodic modulation can be verified.
This complicates the correction of the two observational effects that we
apply to the observed histogram. In order to proceed we choose to {\it
assume} that the samples of the announced planet candidates are complete
up to 40 \ms. Although we have no doubt that this assumption does not
represent accurately the present status of the various studies, it
nevertheless enables us to {\it estimate} the selection effect and
correct for it. Obviously, we excluded from the sample~HD 177830, which
has an amplitude smaller than this arbitrary threshold.

We further assume that there are two phases of the planet candidate
surveys. At the first phase the research groups of Marcy et al. and
Mayor et al. monitored about 200 stars, out of which 8 planet candidates
were found. We assumed this phase is close to completion. The other
phase includes about 1000 additional stars, but, on the other hand, many
more planets are expected to be found in these samples, so the scaling
factor of this phase is not well known. We therefore {\it assumed
arbitrarily} that the new planet candidates came from a sample of 400
stars and averaged and scaled the results of the two phases
accordingly. The results of the calculation are presented in the lower
panel of the figure.

Obviously, all these approximations can not but obscure the frequency of
the planet candidates and the exact shape of their mass
distribution. This is why we choose not to assign any error bars to the
various bins.  The only two goals of the derivation of the histogram are
to estimate the mass distribution {\it slope}, or rather the {\it
direction} of the slope, and the mass distribution boundaries. Although
there are still many unknown factors, we suggest that these two features
of the distribution are already clear. As concluded by many studies
(e.g., Mayor \& Udry 2000) the planet candidates are not part of the
low-mass tail of the secondary mass.  They compose a different
population, well separated on the mass axis, and therefore can be
related to as proper planets. We find that the planet mass distribution
starts with very low values at the 30--100 \MJ\ region, and rises
steeply, even on a logarithmic scale, towards Jupiter and sub-Jupiter
mass range. Stepinski \& Black (2000) got
a different slope for the planet mass distribution, probably because
they took a conservative stand and did not correct for the undetected
systems.

\section{Metallicity}

In this section we turn our attention to the metallicity of the stars
around which the planets have been discovered.  Most of these
stars exhibit metallicity higher than that found in the solar neighborhood
(Gonzalez 1997; Marcy \& Butler 1998; Queloz et al. 2000; Gonzalez 2000;
Butler et al. 2000). Queloz et al. (2000) and Butler et al. (2000)
further pointed out that the host stars to the ``51 peg like'' planets
are particularly metal-rich.  In this section we further study the
dependence of the metallicity on their orbital period, a dependence
plotted in Figure~5.  

Whenever available we have used metallicity derived from spectral
analysis of the stars, mostly from the seminal work of Gonzalez
(2000). Mazeh et al.~(2000) derived the metallicity of
HD~209458. Whenever such an analysis was not available, we have used
photometric metallicity derived by Gim\'{e}nez (2000).  The
metallicities of the stars not considered by Gim\'{e}nez was derived by
us following his prescription, based on the photometry of Hauck \&
Mermilliod (1998) and the calibrations calculated by Crawford (1975) and
Olsen (1984).  Following Gim\'{e}nez (2000), we did not include in the
plot GJ876 and HD177830, for lack of data. Like in Gim\'enez (2000),
HD114762 was excluded because of its extremely low metallicity, and
55~Cnc and 14~Her because of their extremely high $\delta
c_1$. Actually, inclusion of the last three stars would only enhance the
effect suggested here.

\begin{figure}
\plotfiddle{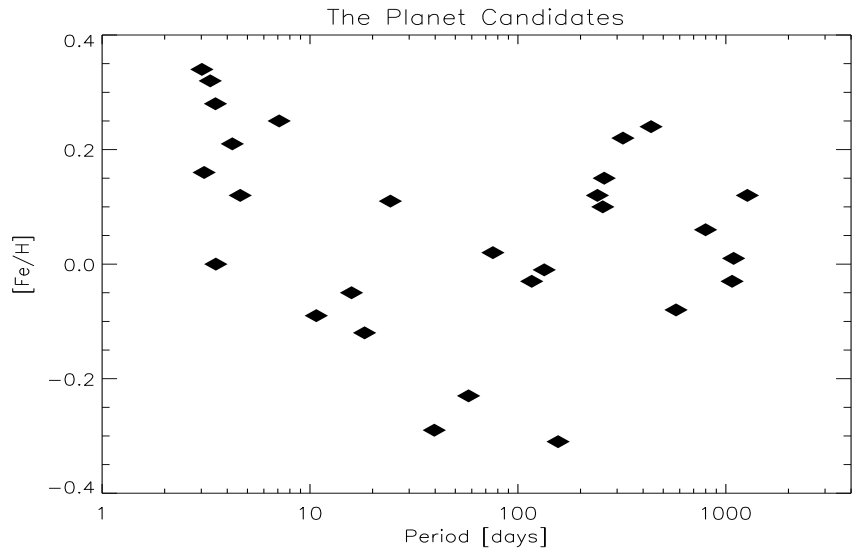}{6.3 truecm}{0}{70}{60}{-165}{-10}
\caption{The metallicity of the parent stars of the extrasolar planets
as a function of their orbital period}
\end{figure}

The figure suggests that the planet candidates with periods shorter than
about 100 days show negative correlation between the
metallicity of their host stars and their orbital period. This
correlation disappears when we consider the planet candidates with
longer periods.  Although the effect is still not established to high
statistical significance, we find this conjecture intriguing, as Section
2 finds a division between the almost circular planets and the eccentric
ones at about the same period.

\section{Conclusion}

The logarithmic mass distribution derived here shows that the planet
candidates are indeed a separate population, probably formed in a
different way than the secondaries in spectroscopic
binaries. Surprisingly the eccentricity and period distribution, with
some restriction, are very much the same.

Furthermore, the two period distributions follow strictly a straight
line. This indicates flat density distributions on a logarithmic scale,
inconsistent with the Duquennoy and Mayor (1991) log-Gaussian
distribution. Interestingly, flat logarithmic distribution 
is the only scale-free distribution, and could be argued to be the
most simple distribution. Maybe the two populations were formed by
two different mechanisms that still have this free-scale feature in
common (Heacox 1999).  

In the last section we present some evidence that the orbital period
of the planets anti-correlates with the metallicity of their host stars.
The aim of this section is to draw the community attention to this {\it
possible} intriguing division, in order to fertilize further
discussion. One interpretation of this {\it possible} effect is that the
planets polluted the stellar atmospheres with heavy atoms from the
early-phase accretion disks when they migrated towards the star. Another
interpretation is that the stars with higher metallicities tend to form
planets more easily. The metallicity can further influence the distance
at which the planet is formed or the distance into which the planet
migrates. In any case, if this effect is confirmed, it is interesting
that it holds only for the close-in planets, for which we do not find
orbits with large eccentricities.
 
Obviously, we need many more planet detections to confirm {\it each} of
the features suggested here . Hopefully, the new high-precision surveys
now in high gear will supply many more planets in the near future,
unraveling the still hidden characteristics of the extrasolar planets.

\acknowledgements We wish to thank D. Goldberg, D. W. Latham, M. Lecar
and R. Noyes for many helpful discussions, and G. Torres for very
illuminating comments on the manuscript.  This research was supported by
grant no. 97-00460 from the United States-Israel Binational Science
Foundation (BSF), Jerusalem, Israel.

\section*{References}
\begin{description}
%\bibitem protect\citename{Abt \& Levy }1976]{al}

\item Burrows, A., Marley, M., Hubbard, W. B., Lunine, J. I.,
 Guillot, T., Saumon, D., Freedman, R., Sudarsky, D., \& Sharp, C. 1997, 
 ApJ, 491, 856

\item Butler, R. P., Vogt, S., Marcy, G.~W., Fischer, D., 
Henry, G. \& Apps, K. 2000, \apj, accepted

\item Carney, B.~W. \& Latham, D. W. 1987, AJ, 92, 116

\item Crawford, D. L. 1975, \aj, 80, 955

\item Duquennoy, A. \& Mayor, M. 1991, A\&A, 248, 485

\item Gim\'{e}nez, A. 2000, \aap, 356, 213

\item Goldberg, D. 2000, Ph.D. thesis, Tel Aviv University

\item Goldberg, D., Mazeh, T., Latham, D.~W., Stefanik, R.~P., Carney,
B.~W., \& Laird, J.~B. 2000, submitted to \aap

\item Goldman. I. \& Mazeh, T. 1991, \apj, 376,260

\item Gonzalez, G. 1997, \mnras, 285, 403

\item Gonzalez, G. 2000  in Disks, Planetesimals and Planets, 
ed. F. Garcon, C. Eiron, D. de Winter, \& T.~J. Mahoney, in press

\item Goodman, J. \& Oh, S.~P. 1997, ApJ, 486, 403

\item Halbwachs, J.-L., Mayor, M., \& Udry, S. 1998,
in Brown Dwarfs and Extrasolar Planets,
ed. R. Rebolo, E.~L. Martin, \& M.~R. Zapaterio Osorio (ASPC) 308

\item Halbwachs, J.-L., Arenou, F., Mayor, M., Udry, S., \& Queloz,
D. 2000, A\&A, 355, 581

\item Hauck, B. \& Mermilliod, M. 1998, \aaps, 334, 221

\item Heacox, W.~D. 1999, ApJ, 526, 928

\item Latham, D.~W., Stefanik, R.~P., Mazeh, T., Torres, G., 
\& Carney, B.~W. 1998, 
in Brown Dwarfs and Extrasolar Planets,
ed. R. Rebolo, E.~L. Martin, \& M.~R. Zapaterio Osorio (ASPC) 178

\item Latham, D.~W., Stefanik, R.~P., Torres, G., Davis, R.~J., Mazeh, T.,
Carney, B.~W., Laird, J.~B., \& Morse, J.~A. 2000, submitted to \aap

\item Lin, D.~N.~C., Papaloizou, J.~C.~B., Terquem, C., Bryden, G. \&
Ida, S. 2000 in Protostars and Planets IV, ed. V.~Mannings, A.~P.~Boss,
S.~S.~Russell (Tucson: University of Arizona Press), 1111

\item Lucas, P.~W. \& Roche, P.~F. 2000, MNRAS, accepted

\item Marcy, G.~W. \& Butler, R.~P 1998, \araa, 36,57

\item Marcy, G. W., Cochran, W. D., \& Mayor, M. 2000 in Protostars and
Planets IV ed. V.~Mannings, A.~P.~Boss, S.~S.~Russell (Tucson:
University of Arizona Press), 1285 (MCM)

\item Mayor, M., Queloz, D., Udry, S., \& Halbwachs, J.-L. 1997, in IAU
Coll.~161, Astronomical and Biochemical Origins and Search for Life in
the Universe ed. C. B. Cosmovici, S. Boyer, \& D. Werthimer 
(Bolognia: Editrice Compositori) 313

\item Mayor, M. \& Udry, S. 2000, in Disks, Planetesimals and Planets, 
ed. F. Garcon, C. Eiron, D. de Winter, \& T.~J. Mahoney, in press

\item Mazeh, T. 1999a, Physics Reports, 311, 317

\item Mazeh, T. 1999b,, in ASP Conf. Ser. 185, IAU Coll. 170, Precise
Stellar Radial Velocities, eds. J.~B. Hearnshaw \& C.~D. Scarfe,  (San
Francisco: ASP), 131

\item Mazeh, T. et al. 2000, ApJ, 532, L55

\item Mazeh, T., Goldberg, D., \& Latham, D. W. 1998, ApJL,
501, L199, (MGL) 

\item Mazeh, T., Mayor, M., \& Latham D.~W. 1996, ApJ, 478, 367

\item Olsen, E.~H. 1984, \aaps, 57, 443

\item Queloz, D., Mayor, M., Weber, L., Bl\'{e}cha, A.,  Burnet, M., 
Confino, B., Naef, D., Pepe, F., Santos, N., \& Udry, S. 2000 \aap, 354, 99

\item Schneider, J. 2000, in Extrasolar Planets Encyclopaedia 
http://www.obspm.fr/planets

\item Stepinski, T. F. \& Black, D. C. 2000, in IAU Symp. 200, Birth and
Evolution of Binary Stars, ed. B. Reipurth \& H. Zinnecker (Potsdam) 167

\item Trilling, D.~E. 2000, ApJ, 537, L61

\item Weidenschilling, S.~J. \& Marzari, F. 1996, Nature, 384, 619

\item  Zahn, J.-P., \& Bouchet, L.  1989, A\&A, 223, 112
 
\end{description}

\end{document}